\documentclass[twocolumn,plb,aps,groupedaddress,nopacs,preprintnumbers,floats,floatfix]{revtex4}

\usepackage[pdftex]{graphicx}
\usepackage{natbib}
\usepackage{amsmath,url}
\usepackage[alpha,z,expert]{tmsfnts}
\usepackage{color}
\usepackage[normalem]{ulem}
\usepackage{booktabs}
\definecolor{BurntOrange}{rgb}{0.81, 0.32, 0.0}
\definecolor{DarkGreen}{rgb}{0.0, 0.5, 0.0}
\definecolor{DarkRed}{rgb}{0.5, 0.0, 0.0}
\definecolor{DarkBlue}{rgb}{0.0, 0.0, 0.5}
\definecolor{purple}{rgb}{0.5, 0.0, 0.5}
\definecolor{red}{rgb}{1, 0.0, 0.0}
\definecolor{green}{rgb}{0, 1.0, 0.0}

\def\3he{$^3{\rm He}$}

\begin{document}
\pagestyle{plain}
\pagenumbering{arabic}


\title{White Paper: Radio Emission and Polarization 
Properties of Galaxy Clusters with VLASS} 

\author{
Tracy Clarke (NRL),
Tony Mroczkowski (NRC Postdoc),
Shea Brown (U. Iowa), 
Gianfranco Brunetti (INAF),
Rossella Cassano (INAF), 
Daniele Dallacasa (INAF),
Luigina Feretti (INAF),
Simona Giacintucci (UMD), 
Gabriele Giovannini (Unibo),
Federica Govoni (INAF),
Maxim Markevitch (GSFC),
Matteo Murgia (INAF),
Lawrence Rudnick (UMN),
Anna Scaife (Southampton),
Valentina Vacca (INAF),
Tiziana Venturi (INAF), 
Reinout van Weeren (CfA)}

\maketitle


\section{Executive Summary}\label{abs}

We report here on a broad range of forefront science goals 
relating to the physics of galaxy clusters that could be addressed by the
upcoming Very Large Array Sky Survey (VLASS).  Based on these science goals
and the complementarity VLASS will have with ongoing and completed surveys,
we discuss observation strategies and provide recommendations for
the bands and configurations with the most potential for scientific 
return for the subset of the galaxy cluster community interested
in diffuse, non-thermal emission.

The VLASS could provide a major contribution in three key areas
of the physics of galaxy clusters:

\begin{itemize}

 \item {\bf The active galactic nucleus (AGN) population in galaxy clusters 
	 and the impact of AGN feedback on the intra-cluster medium (ICM).} 
	 Extended radio galactic structures such as narrow and wide 
	 angle tails (NATs and WATs) trace ICM weather, while the
	 radio emission associated with the cluster dominant galaxies 
	 is key in the study of the ICM/AGN feedback.

 \item  {\bf The origin and evolution of diffuse cluster radio sources}.
	 Radio halos, minihalos and relics are direct signposts of the dynamical
	  state of the ICM, and may probe the role of shocks and turbulence in 
	  the formation and evolution of large scale structures in the Universe.

 \item {\bf The origin and role of magnetic fields in the turbulent environs of
	 the ICM and in large scale structures.} Cluster magnetic fields are 
	 known to reach levels of several $\mu$Gauss, but their evolution and 
	 growth is not fully understood.

\end{itemize}

The evolution and interplay of baryons and magnetic fields in clusters, 
the galaxies within clusters, and large scale structure in general have 
been identified as key studies for this decade \citep{Kravtsov2009,Myers2009}.
Toward this end, we recommend VLASS wideband continuum survey strategies that 
attain high surface brightness sensitivity on scales from a few arcseconds and
recover scales out to several arcminutes.  {\bf The bands and configurations
for large scale diffuse cluster emission include S Band (2--4~GHz) in VLA D 
(and possibly C) Configuration, L Band (1--2~GHz) in C Configuration, and P band 
(230--470~MHz) in B Configuration.}  Further, in order to probe cluster weather and 
feedback from AGN, as well as to subtract sources of contamination from the 
diffuse large scale emission, we recommend complementary observations in 
S Band, B Configuration.


\section{Introduction: Galaxy Clusters}\label{intro}

Clusters of galaxies are the largest gravitationally bound systems in
the Universe, and are dominated by dark matter ($\sim$80\%). Only a 
tiny fraction of a cluster's mass is in the form of stars in galaxies 
($\sim$3--5\%), while the rest ($\sim$15--17\%) comprises the intra-cluster
medium (ICM), which is diffuse hot (10$^{7-8}$~K) gas detected in X-ray observations 
by its thermal bremsstrahlung and highly-ionized line emission. 
A large improvement in the present knowledge of the astrophysics in galaxy 
clusters has been obtained in recent years from the study of the ICM through 
the combination of X-ray and radio observations.  

Clusters form by hierarchical structure formation processes. In
this scenario, smaller units (galaxies, groups and small clusters)
formed first and merged under gravitational pull to larger and larger
units in the course of time. Cluster mergers are the primary mechanism by
which clusters and superclusters are assembled.  Denser regions form a filamentary
structure in the Universe, and clusters form at the intersections of
these filaments. Major cluster mergers are among the most energetic events
in the Universe \citep{Sarazin02}.  During mergers, shocks are driven
into the ICM, with the subsequent generation of turbulence. The merger
activity appears to be continuing at the present time and, along with
feedback from AGNs and star formation, explains the
relative abundance of substructure and temperature gradients detected
in clusters of galaxies by optical and X-ray observations.

Clusters can reach a relaxed, nearly virialized state characterized by a 
giant galaxy at the center and enhanced X-ray surface brightness peak in 
the core.
The hot gas in the center has a high density, which implies short
radiative cooling times; therefore energy losses due to X-ray emission
are dramatic and produce a temperature drop towards the
center. Relaxed clusters were then classified as ``cooling flow''
clusters \citep{Fabian94}.  This model was the subject of much debate,
when XMM-Newton spectral results failed to confirm the lines and
features expected as a product of a steady state cooling flow
\citep{Peterson01,Peterson06}.  The classical cooling flow model has
finally been replaced by the ``cool-core'' paradigm where some
heating source offsets the catastrophic cooling expected in the cooling
flow model.  At present, it is widely accepted that the source of
heating in cool-core clusters is the AGN activity of the brightest
cluster galaxy at the center \citep[see][for recent
  reviews]{Mcnamara07,Bohringer10}.

Galaxy clusters are spectacular systems in the radio band.  Obvious
radio sources are the individual galaxies, whose emission has been
observed in recent decades with sensitive radio telescopes. It often
extends well beyond the galaxy optical boundaries, and hence it is
expected that the radio emitting regions interact with the ICM.  This
interaction is indeed observed in tailed radio galaxies, and radio
sources filling X-ray cavities at the center of cool-core clusters
\citep[e.g.][]{Mcnamara07,Feretti08}.

More puzzling are diffuse extended radio sources, which cannot be
obviously ascribed to individual galaxies, but are instead associated
with the ICM.  This radio emission represents a striking feature of
clusters, since it demonstrates that the thermal ICM plasma is mixed
with non-thermal components.  Diffuse sources are typically
observationally identified as halos, relics, or minihalos
according to their size, location in the cluster,
polarization properties, and the dynamical state of the host cluster (merging
or cool-core) \citep{Feretti96}. These systems all require the presence of large-scale
magnetic fields and a population of relativistic electrons spread over
100's of kpc to Mpc scales throughout the cluster volume. Further
demonstration of the existence of magnetic fields in the ICM is
obtained by studies of the Faraday rotation of polarized radio
galaxies lying in the background or embedded within the magnetized
intra-cluster medium.

Non-thermal components are important for a comprehensive physical
description of the intra-cluster medium in galaxy clusters 
\citep{Subramanian2006,Schekochihin2010,Brunetti2011}, and play a
major role in the evolution of large-scale structures in the Universe.
The discovery of diffuse cluster radio emission presents an
important step in the understanding of the physical processes in
clusters of galaxies \citep[see][for a recent review]{Feretti12}.
Diffuse synchrotron sources are sensitive to the turbulence and shock
structures of large-scale environments and provide essential
complements to studies at other wavebands as well as unique physics
not probed by any other wavelength regime. Studies in the radio domain
will fill essential gaps in both cluster astrophysics and in the
growth of structure in the Universe, especially where the signatures
of shocks and turbulence, or even the thermal plasma itself, may be
otherwise undetectable.

In the following sections of this white paper, we make the case for
the key scientific focus areas for a VLASS survey sensitive to
extended non-thermal emission from clusters (\S~\ref{science}). We
provide a brief overview of relevant multi-wavelength data sets important 
for cluster studies in \S~\ref{complements}, and discuss available and
upcoming radio surveys in the context of their importance for clusters and
integration to the VLASS in \S~\ref{surveys}. Section ~\ref{requests}
provides the specifics of the survey configuration possibilities and
their applicability to cluster science, and it highlights our recommended
survey strategies for maximizing scientific return for cluster astrophysics. 


\section{Extended Non-Thermal Emission from Galaxy Clusters}\label{science}

\subsection{AGN and the Environment: Lifecycle and Feedback}\label{agn}

Radio emission from individual cluster galaxies provides invaluable
information on the formation and evolution of the hosting structures,
mainly along three main branches: (i) AGN activity and AGN/ICM
feedback in the central cluster regions; (ii) distorted wide angle
tail (WAT) and narrow angle tail (NAT) radio galaxies as probes of the
cluster dynamical state and cluster weather, as well as signposts of
high-z clusters; (iii) the role of cluster dynamics on the radio
emission of individual galaxies through the radio luminosity function.

\begin{figure}[tbh!]
\footnotesize	
 \begin{centering}
  \includegraphics[width=3.25in]{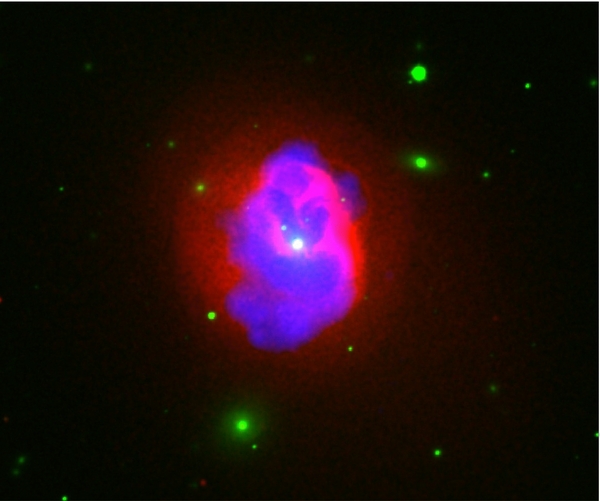}
  \caption{\small AGN feedback in Abell 2052 ($z=0.035$), from \cite{Blanton2011}.
  }\label{fig:AGN}
 \end{centering}
\end{figure}

\subsubsection{AGN Feedback}

Some galaxy clusters known as cool core clusters display
highly peaked X-ray emission and are generally hosts to cluster-center
radio galaxies (CCRGs). The probability of hosting a CCRG increases
from 45\% for non-cool cores to 67\% for weak cool core and up to
100\% for strong cool core clusters \citep{Mittal09}. These powerful
active galactic nuclei (AGN) are currently considered one of the
favored sources of energy input into the ICM to offset radiative
cooling which, left unopposed, would eventually develop into a runaway
cooling flow \citep{Fabian94}. AGN-induced feedback is also considered
a key component in shaping the luminosity function of the host galaxy
\citep{Croton06}, it may set the upper limit to host galaxy masses,
and it is expected to contribute to pre-heating of the ICM
\citep{Dubois11}.

Deep X-ray observations toward cool-core systems discovered X-ray
cavities filled with synchrotron plasma visible at GHz
frequencies. The most well-studied AGN cavity systems are Perseus
\citep{Fabian11}, Abell 2052 (\cite{Blanton2011}; see Figure \ref{fig:AGN}), 
Hydra A \citep{Wise07}, Virgo A \citep{Forman05}, and NGC 5813
\citep{Randall11}. These observations clearly indicate that the
radio galaxy is significantly impacting the ICM, inflating cavities in
the thermal gas and driving weak shocks and sound waves through the
ICM. These bubbles are expected to detach and buoyantly rise through
the ICM after the central AGN activity decreases. In addition to the
energy injection, these rising bubbles provide a means of
seeding the ICM with magnetic fields and relativistic particles.

X-ray observations are only able to easily detect small cavities near
the plane of the sky at relatively small distances from the
cluster core. Larger cavities at large distances from the core, where
the ICM is more diffuse, as well as those at small angles from the 
line of sight do not provide sufficient contrast for detection in even 
moderately deep X-ray observations \citep{Ensslin02}.  
On the other hand, radio observations of CCRGs provide methods to place 
observational limits on the energy injected into the ICM by AGN feedback by
tracing the complete kinetic feedback history of the ICM over multiple 
AGN outburst cycles. As a matter of fact, a number of central galaxies 
in clusters and groups are characterised by extended (10--100~kpc) diffuse and faint aged 
synchrotron emission, best detectable at frequencies of few hundred MHz, with an active
radio nucleus, whose spectrum clearly shows ongoing activity. Despite
the many uncertainties, the study of the radio spectrum in the aged
and active components provides reliable information on the cycles of
activity of CCRGs, and the total energy output delivered into the
cluster ICM throughout the cluster lifetime (e.g. \cite{Giacintucci12}).
%


\begin{figure}[tbh!]
\footnotesize
 \begin{centering}
  \includegraphics[width=3.25in]{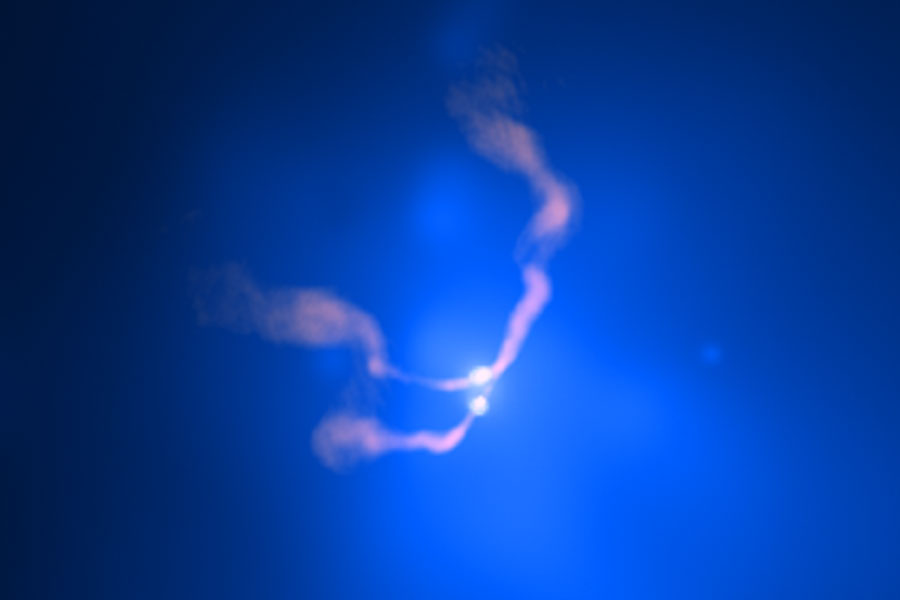}
  \caption{\small Wide Angle Tails of the active galaxy 3C\,75, located in the cluster 
  environment of Abell 400. Figure is from \cite{Hudson2006}.
  }\label{fig:WAT}
 \end{centering}
\end{figure}

\subsubsection{Cluster Weather and High-$z$ Clusters}\label{weather}

WAT and NAT radio galaxies are the most spectacular examples of radio
emission from elliptical galaxies (see Figure \ref{fig:WAT}). 
Their shape is the signature of
galaxy cluster membership, and is explained as due to the combination
of motion of the hosting galaxy within the cluster and intra-cluster
bulk motion \citep{Blanton00,Blanton01,Wing11}. Spectral studies along
the tails provide estimates of the age of the radio plasma, which in
turn can be used to infer the galaxy velocity within the cluster, and
information on the dynamical state of the cluster.  Due to their
unique association with dense environments, WATs and NATs can be used
to easily identify high-$z$ galaxy clusters. Note that high-$z$ clusters
require significant observational efforts for detection in the X-ray and
optical bands, while they are fairly accessible with high resolution
(arcsecond scale) radio observations \citep{Giacintucci09, Mao10}.
In this way, wide radio surveys that discover WATs and NATs nicely 
complement the ongoing surveys that exploit the redshift-independent 
surface brightness of the Sunyaev-Zel'dovich (SZ) effect, which is now being 
used to located previously-unknown clusters at high-$z$ (see \S \ref{complements}) . 

Radio galaxy bending, forming WATs, NATs, or even distorted FRII radio galaxies, 
is sensitive to the pressure gradients and the relative motions through the ICM.  
While simple swept-back radio structures could be due to galaxy motions, more 
complicated morphologies require `weather' in the cluster, as expected by the 
ongoing accretion of material along filaments.  This weather is normally 
undetectable in X-ray or SZ observatations, and is only visible in those regimes
when there is a significant contact discontinuity (i.e. a shock or cold front).   
However, radio galaxy distortions can easily pick out
the transonic or even subsonic motions that persist even when the cluster may
appear relaxed in the X-rays.  Such studies will require large samples to 
eliminate structures that could simply be due to cluster motion, and to look for
the correlations between weather indicators and evolutionary state as indicated 
from X-ray or SZ images.
We also highlight here the strong complementarity of these
observations with X-ray observations with ASTRO-H, and with Athena+ in 
about 15 years, that are aimed at the measure of turbulent motions
from the study of metal lines in the X-ray spectra of galaxy clusters 
\citep{Takahashi2010}.

One critical discontinuity that has so far eluded detection is the accretion
shock, where the infalling material is first heated as it accretes onto the 
cluster.  Simulations show that the accretion shock for massive clusters likely
exists at several Mpc from the cluster center, outside the virial radius, 
but the thermal gas is currently too diffuse to be detectable directly
in X-ray observations.  
Radio galaxies have the potential, however, to statistically map out this 
critical transition region, because a galaxy's infall through the shock region
will be relatively unaffected, while the tails will be distrupted as they
interact with the shocked ICM.  
Rudnick and students (2013, in preparation) have used the bending 
classifications of FIRST sources by \citet{Wing11} to examine the 
prevalance of bending as a function of distance from the cluster center.  
While a cluster-associated population of WATs/NATs can be statistically detected out to at 
least 10~Mpc, the numbers are so far only high enough to detect tail disruption
within the inner $\sim1$~Mpc.   With larger samples enabled by probing $>1.5\times$ 
deeper than FIRST and recovering the larger scales appropriate to WATs/NATs
(see Table~\ref{table:features}), this pioneering work may be extended 
to larger cluster radii, with the potential of identifying the otherwise 
invisible accretion shock.


\subsubsection{AGN--ICM Connection and Scaling}\label{lumfunc}

A longstanding question in our understanding of the trigger of nuclear
radio activity in AGN is whether the environment plays a role, and how. The
radio luminosity function for galaxies in different environments is the primary
statistical tool for investigation in this area. The wealth of data available
from public archives in the X-ray and optical bands now allows us to
address the role of cluster dynamics (i.e.\ merger versus relaxed systems) 
on the AGN activity. This research has been impossible until recently, and has
been limited mainly by the statistical information in available radio
survey data.

A new L or S band survey of intermediate depth ($\sim$70--90~$\mu$Jy/bm)
would considerably increase this statistical information. 
An improved sensitivity of a factor of 5--7$\times$ over 
NRAO VLA Sky Survey (NVSS; \cite{Condon1998}) would bring the 
radio power limit down to $\sim 10^{20}$~W~Hz$^{-1}$
in the local Universe (i.e. $z=0.02$), thus allowing exploration of the faint
end of the radio luminosity function and separation of the AGN and starburst
contribution. At higher redshifts (e.g. $z=0.2$--0.5), the radio power
limit would be of the order of $\sim 10^{23}$~W~Hz$^{-1}$, still within the FRI
range. This will allow the study of both evolutionary and environmental effects
in the population of low and high luminosity radio galaxies (FRI and FRII).


\begin{figure}[tbh!]
 \begin{centering}
  \includegraphics[width=3.25in]{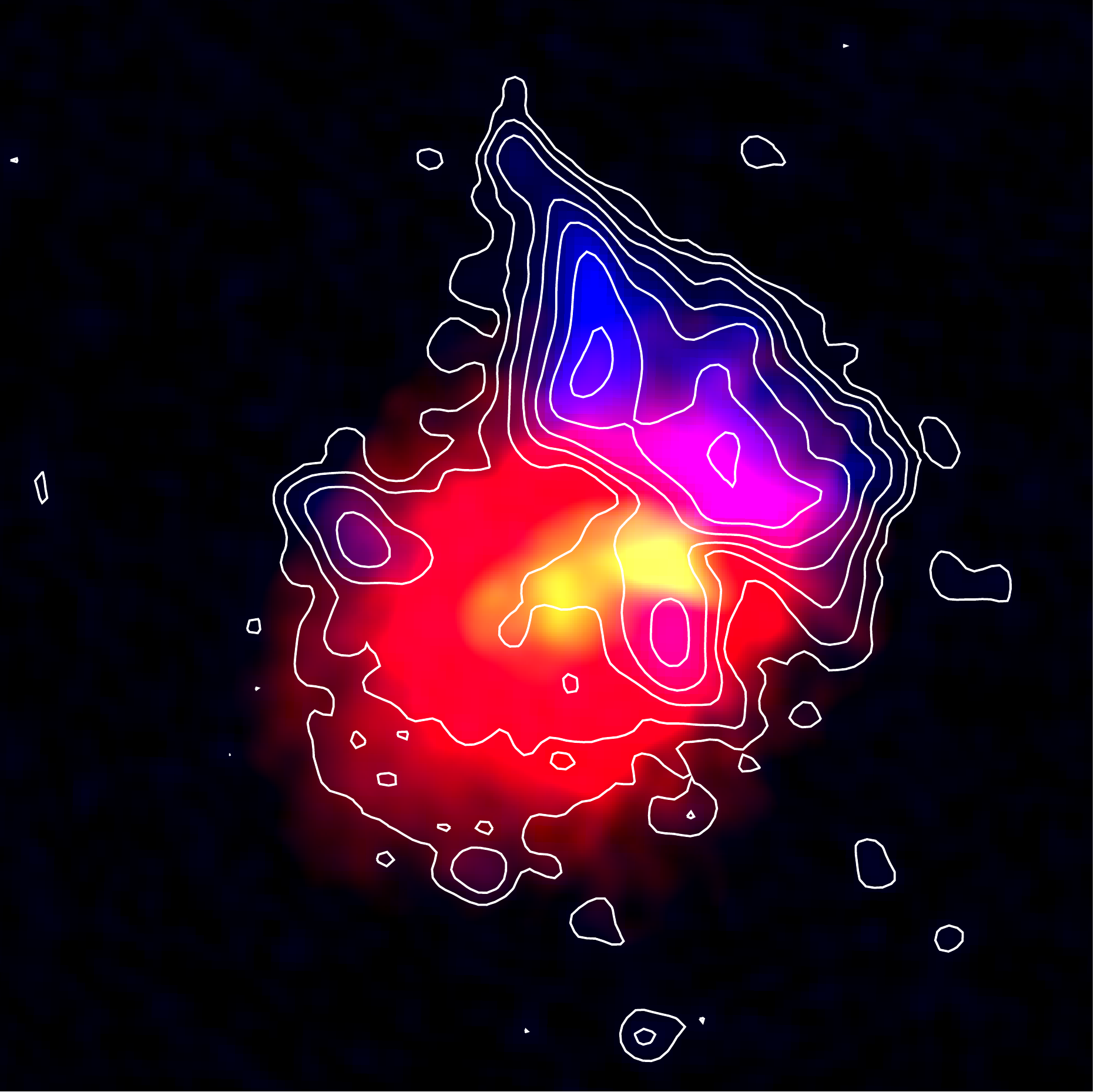}
  \caption{\small The merging cluster Abell 2256 ($z=0.0594$) shown in X-rays in red 
	  and green image and radio (1.4 GHz) in blue and white contours. 
	  The X-ray emission shows two clear peaks, indicating a significant 
	  on-going merger, and the radio shows the presence of a Mpc scale 
	  central radio halo as well as Mpc radio relics.
	  Figure is from \citet{Clarke2006}.
  }\label{fig:relichalo}
 \end{centering}
\end{figure}

\subsection{Dynamically Complex Clusters: Halos and Relics}\label{halorelics}

In about 70 clusters, diffuse, Mpc-scale radio emission has been found,
implying the presence of relativistic particles and magnetic
fields. These giant halos and relics are only found in dynamically
disturbed clusters that show clear evidence for undergoing one or
multiple merger events (see Figure \ref{fig:relichalo}). 
Halos and relics generally have steep
synchrotron spectra ($\alpha \lesssim -1$, where flux density $S_\nu \propto \nu^\alpha$), 
large physical extents ($\sim$1--1.5~Mpc), and 1.4~GHz radio powers in the range of
$10^{23-26}$~W~Hz$^{-1}$ \citep[see][for a recent review]{Feretti12}. 
An important aspect of the study of halos and relics concerns the origin of the
relativistic particles and magnetic fields. The lifetime of the
synchrotron radiating electrons is much shorter than the diffusion
time necessary to fill the volume these sources occupy. Therefore a
form of in-situ particle (re)acceleration is required.

\begin{figure}[tbh!]
\footnotesize
 \begin{centering}
  \includegraphics[width=3.25in]{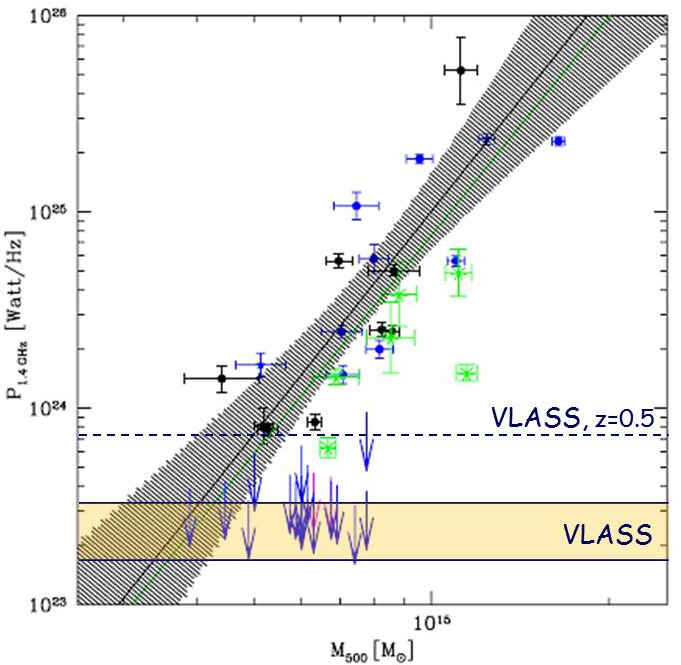}
  \includegraphics[width=3.25in]{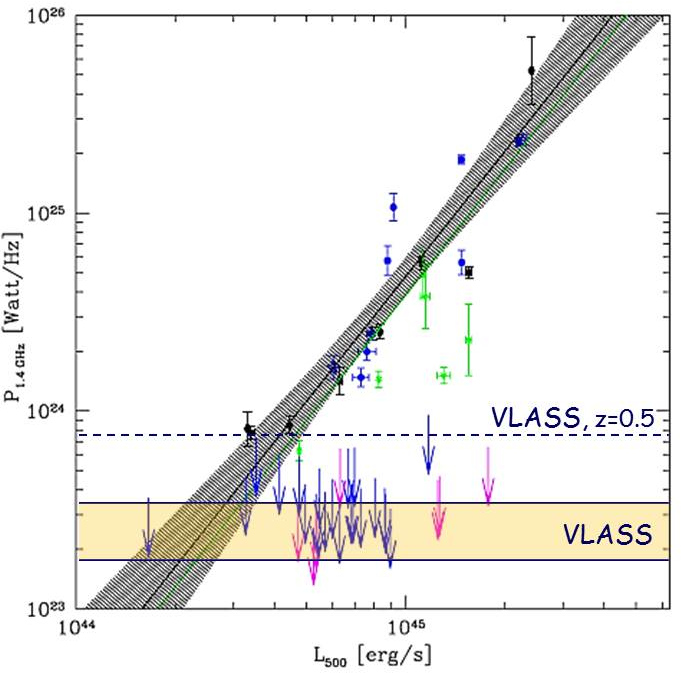}
  \caption{\small Radio halo power and VLASS sensitivity to giant halos as
	  a function of  cluster mass and X-ray luminosity, using the approach in 
	  \cite{Cassano2010,Cassano2012} and the information in 
	  Table~\ref{table:configs}.
	  \cite{Cassano2013} estimated the sensitivity of surveys to giant 
	  halos by taking into account the brightness distributions of radio 
	  halos. The yellow region marks the sensitivity of VLASS in the redshift 
	  range $z=0.2$--0.3, equivalent to the redshift range of the 
	  GMRT halos (for details see \cite{Cassano2013}). 
	  Based on these calculations, a survey in L or S Band would not 
	  differ significantly in its ability to detect new halos.
	  Consequently, we report the case of L Band in C Config with 
	  RMS = 1.5$\times$ the confusion (Table \ref{table:configs}).  
	  The dashed line is the sensitivity at $z=0.5$.
  }\label{fig:scalings}
 \end{centering}
\end{figure}

Giant radio halos are centrally located and are distributed
co-spatially with the X-ray emitting intra-cluster medium (ICM).  For
radio halos there is a correlation between the X-ray luminosity
$L_{\rm{X}}$ (i.e., cluster mass) and radio power $P_{1.4\rm{GHz}}$
(e.g.\ \cite{Liang2000}; see Figure \ref{fig:scalings}). 
The fraction of clusters hosting radio halos is still uncertain.
Current statistical analyses suggest roughly 30--40\% of the most X-ray luminous
systems at low/intermediate redshifts ($z \lesssim 0.4$) host radio halos
\citep{Giovannini2009, Venturi2007, Venturi2008}, while there is 
evidence that this fraction decreases at lower X-ray luminosity
and mass \citep{Cassano2008,Cassano2013}. 
A complementary approach using massive, SZ-selected clusters from the {\it Planck} catalog 
indicates the fraction of SZ-selected clusters hosting radio halos is much 
higher ($\sim$80\%) \cite{Sommer2013}.

Radio halos have been explained by turbulence injected by recent merger events, 
which re-accelerates relativistic particles \citep{Brunetti01,Petrosian01}.
In a competing model, the energetic electrons are secondary products
of proton-proton collisions (e.g., \cite{Dennison1980}); 
however, important progress has been made that disfavors this second 
scenario. This includes (i) the
discovery of ultra-steep spectrum radio halos that likely cannot be
explained by secondary models \citep{Brunetti2008}, (ii) a clear
relation of halo host clusters with merger signatures
\citep{Cassano2010,Brunetti2007}, and (iii) the lack of $\gamma$--ray
detection by {\it Fermi} \citep{Ackermann2010,Brunetti2012,Zandanel2013}.  
This progress has been achieved by carrying
out observations of representative cluster samples
\citep{Venturi2007,Venturi2008,Kale2013}, as well as by targeted deep
observations \citep[e.g.,][]{Macario2011,BrownRudnick2011,Feretti12}.

Unlike halos, relics (again, see Figure \ref{fig:relichalo}) are mostly found 
in the outskirts of clusters and show a high degree of polarization. 
Like halos, the fraction of clusters with radio relics has been found to 
increase with $L_{\rm{X}}$, to about 30\% for the most massive clusters 
\citep{vanWeeren2011, Nuza2012}. 
Particularly interesting are the class of double relics, with the two relics 
located symmetrically on opposite sides of the cluster center tracing 
the two outgoing merger shocks
(e.g. \cite{vanWeeren2010,vanWeeren2011b,vanWeeren2012, Bonafede2009, 
Bonafede2012, Kale2012, Bagchi2011}). 
These relics often show spectral index gradients (rather than being
described by a single or broken power law), as expected if they 
trace merger shocks traveling outwards.  Long ($\gtrsim 500$~kpc) relics have
been explained by particles directly (re)accelerated at shocks by the diffusive
shock acceleration  mechanism (DSA) in a first order Fermi process. However, 
cluster merger shocks typically have low Mach numbers ($\mathcal{M}\sim$1--3, e.g., 
\cite{Markevitch2002, Russell2010, Akamatsu2013, Ogrean2013}) and the efficiency 
with which such shocks can accelerate particles is unknown. Therefore, 
re-acceleration of pre-accelerated electrons in the ICM might be required 
to explain the observed brightness of relics, since re-acceleration is a more 
efficient mechanism for weak shocks 
(e.g., \cite{Markevitch2005, Giacintucci2008, KangRyu2011, Kang2012, KangRyu2013}). 
Such preexisting relativistic electrons could be fossil
radio plasma deposited in the ICM by active radio galaxies.
The shock re-acceleration model also explains
why in some cases clear shocks are observed in the X-rays while no radio relics
are present \citep{Russell2011}.

The properties of radio halos and relics provide direct 
proof of the presence of relativistic electrons and magnetic fields
within the cluster volume.  Hence, studies of these features provide a
unique opportunity to probe the strength and structure of the magnetic
field on Mpc scales \cite[e.g.][]{Vacca10}. Of equal importance, the
location and properties of these diffuse non-thermal sources can be
related to cluster characteristics derived from optical and X-ray
observations, and are tightly connected to the cluster's evolutionary
history. In particular radio halos and relics are always located in
clusters showing merging processes even if not all merging clusters
show the presence of a diffuse radio emission. The VLASS survey could
address the remaining gaps in the particle acceleration mechanisms
through deep observations of a few of these complex merging
environments as well as through statistics from a large survey
providing information on the relativistic particle and magnetic field
content across a range of cluster environments.

In Figure \ref{fig:scalings}, we show predictions for radio halo detection in
an L or S Band survey that approaches $1.5\times$ the confusion limit, 
corresponsing to a sensitivity of 16~$\mu$Jy/bm for L Band in C Config
or 21~$\mu$Jy/bm for S Band in D Config (see Table~\ref{table:configs}).
At these sensitivities, VLASS would be complete to 
$M_{500}\approx3\times10^{14}~\rm M_\odot$ at $z\sim0.3$, a factor of 2
lower than current surveys with the GMRT, which are only complete 
at the $\sim50\%$ level percent for $M_{500}>6\times10^{14}~\rm M_\odot$.
A deep VLASS would therefore chart unexplored territory in radio halo statistics.


\begin{figure}[tbh!]
\footnotesize
 \begin{centering}
  \includegraphics[width=3.25in]{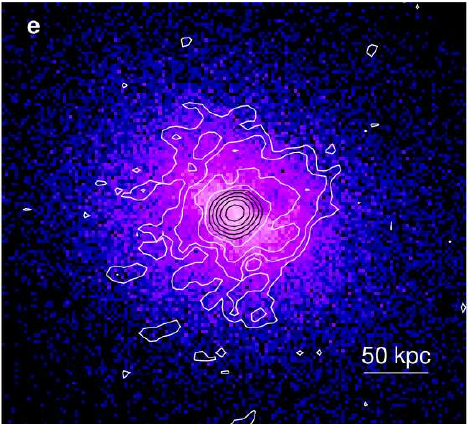}
  \caption{\small VLA 1.4~GHz observations (contours) of RX J1532.9+3021 ($z=0.36$) overlaid 
  on the {\it Chandra} X-ray image.  The extended minihalo is denoted by white contours
  to separate it visually from the central AGN.  Figure is from \citet{Giacintucci2014}.
  }\label{fig:minihalo}
 \end{centering}
\end{figure}

\subsection{Dynamically Relaxed Clusters and the Radio Minihalos in Their Cores}\label{minihalos}

Some dynamically-relaxed clusters are known to host centrally-located, diffuse 
synchrotron radio emission in their cores, that typically fills the central 
cooling region ($r\sim 50-300$ kpc; see Figure \ref{fig:minihalo}). 
These extended radio sources -- called minihalos -- typically have low surface 
brightnesses ($> 2 \mu$Jy arcsec$^{-2}$) and steep radio spectra ($\alpha < -1$) 
\cite{Giacintucci2014}. 
Their emission encompasses the often-present central radio galaxy but extends to 
greater radii. 
An example is the minihalo in Perseus cluster, whose emission spans 
much larger radii than the inner $r\sim 30$ kpc region occupied by the prominent
X-ray cavities and lobes of 3C\,84. Only 15 clusters have confirmed minihalos,
all of which are hot and massive systems with very X-ray luminous cool cores 
(\citet{Giacintucci2014} and references therein).

The origin of these minihalos is still unclear. 
One attractive possibility is that the radio emission arises 
from pre-existing, aged relativistic electrons (for instance, from past activity
of the central radio galaxy and/or hadronic collisions) that are being 
re-accelerated to ultra-relativistic energies by turbulence in the ICM 
(\cite{Gitti2002}).  
Sloshing of dense gas in the cluster cores, revealed by arc-shaped
cold fronts often observed in high-resolution X-ray images of cool-core clusters, 
can amplify the magnetic field 
in the core and generate turbulence that may be strong enough to re-accelerate 
low-$\gamma$ electrons to $\gamma\sim10^4$. 
The combined effect is diffuse radio emission with morphology, radio power, and 
spectral index consistent with the observed minihalos (\cite{ZuHone2013}). 
A spatial connection of the minihalo radio emission and the X-ray sloshing cold fronts 
has indeed been observed (\cite{Mazzotta2008, ZuHone2013, Giacintucci2014}), supporting 
the hypothesis that radio-emitting electrons are re-accelerated by sloshing. 
This opens an interesting possibility of studying low-level turbulence in the 
cores of relaxed clusters and -- in conjunction with the forthcoming X-ray 
probes of ICM turbulence (Astro-H and possibly other future missions) -- the 
efficiency of cosmic ray acceleration by MHD turbulence. Despite recent theoretical
effort, the study of minihalos has been severely limited by their small number; only
a fraction of cool-core clusters are known to host minihalos. 
A much larger sample is needed to determine their occurrence in clusters of various 
masses and cool-core types in order to investigate their origin and the relevant
aspects of the ICM physics.


\subsection{Whither the WHIM: Where is the Warm Hot Intergalactic Medium?}\label{whim}

\begin{figure}[tbh!]
\footnotesize
 \begin{centering}
  \includegraphics[width=3.25in]{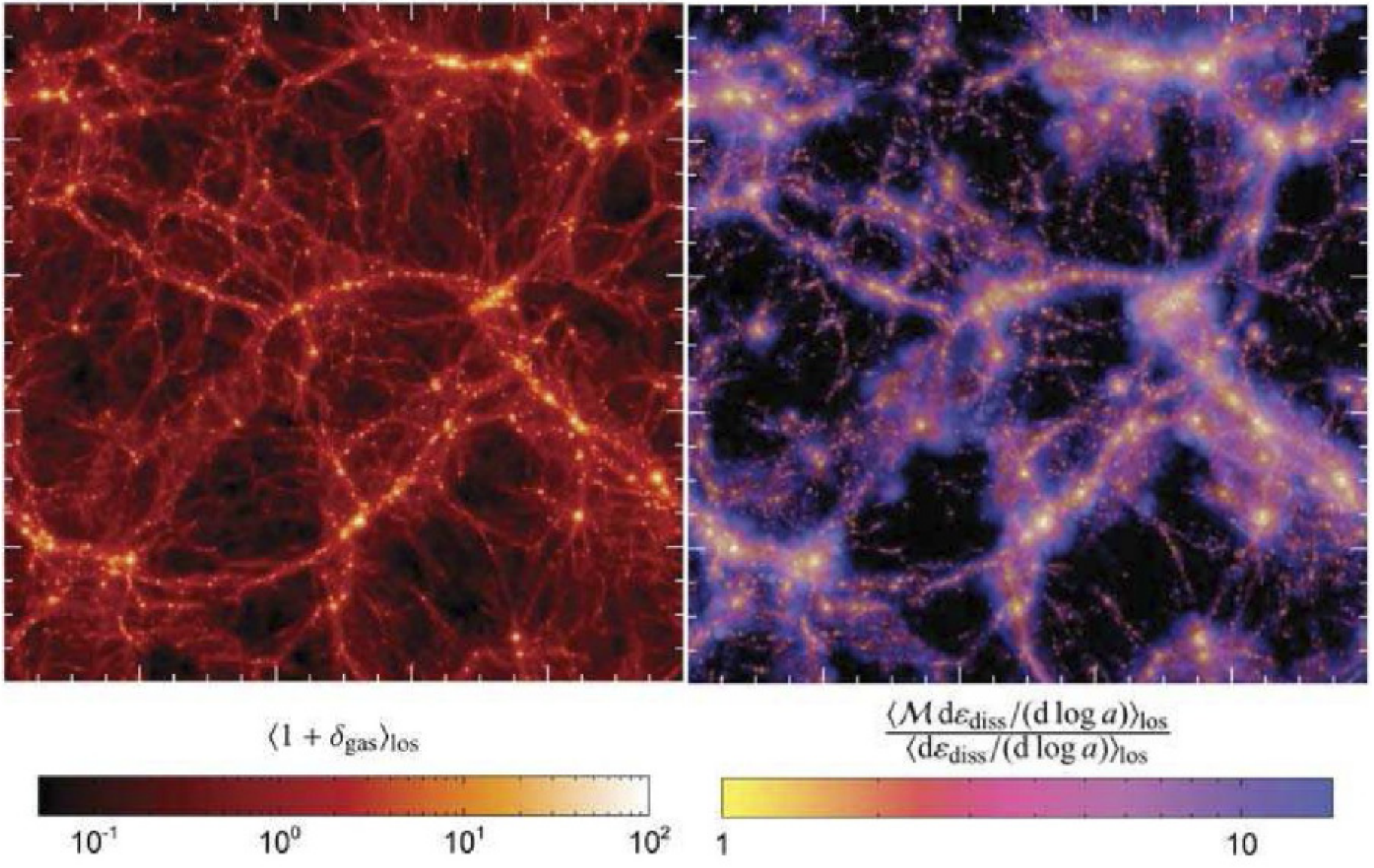}
  \caption{\small  {\bf Left:} gas density from the cosmological simulation 
	  reported in \citet{Pfrommer2006}.  
	  {\bf Right:} Mach number of cosmic shocks in the WHIM, weighted by the
	  energy injected into cosmic-rays.  Figure is from \citet{Pfrommer2006}, 
	  with caption from \citet{Brown2011}.
  }\label{fig:whim}
 \end{centering}
\end{figure}

Approximately 50\% of baryons in the Universe are currently unobserved,
based on the estimates of $\Omega_{b}$ from nucleosynthesis and WMAP
\citep{Bregman2007}. Simulations suggest that the collapsing diffuse
intergalactic medium (IGM) was shock-heated and now resides in filaments
as the warm-hot intergalactic medium (WHIM) with temperature T$\sim$10$^{5-7}$~K 
(see Figure~\ref{fig:whim}).  Due its temperature, it is practically invisible 
at X-ray and optical wavelengths (\cite{CenOstriker1999, Dave2001}). 
However, shocks and turbulence from infall into
and along the filamentary structures between clusters are now widely
expected to generate relativistic plasmas which track the distribution of
the WHIM (the ``Synchrotron Cosmic-Web", 
\cite{Keshet2004, Pfrommer2006, Ryu2008, Skillman2008, Araya-Melo2012}).
If such features are detected in the radio, they can be used to set limits
on the (invisible) pressure of the thermal gas, delineate shock
structures, and illuminate large scale magnetic fields.

However, observational confirmation of this paradigm remains to be seen.
Obstacles include very low intrinsic surface brightness
($<\mu$Jy~arcmin$^{-2}$ levels; \cite{Pfrommer2008, Battaglia2009}), 
large spatial scales ($>$Mpc), and numerous sources of confusion
(see \cite{Brown2011} for a review). The most promising avenue for the VLASS to
detect synchrotron emission associated with the WHIM is to detect shock
structures in the filaments surrounding massive clusters of galaxies,
including the powerful virial shocks. At frequencies greater than 1~GHz,
radio observations suffer from lower intrinsic source surface brightness
and insensitivity to large angular scales. However, polarization
observations will provide the biggest gains when working above 1~GHz,
especially when searching for these shock-structures. Structure formation
shocks onto and along filaments are predicted to be narrow,
flat spectrum ($\alpha>-1$), and highly polarized (e.g., 
\cite{Ryu2008, Skillman2011}). Therefore, both direct and statistical detection of
these shocks can ideally be performed at $\nu > 1$~GHz (e.g. S Band). At the levels
proposed by the VLASS survey in S-Band, only shocks with enhanced
brightness due to, e.g., interaction with extended radio galaxies, would
be detected directly. Stacking clusters with similar mass and distances,
however, could reveal emission due to virial shocks if present
 \citep{Brown2011b}.

 
\subsection{Faraday Rotation Measure (FRM) Synthesis/Polarization}\label{pol}

Magnetic fields can be found in almost every place in the Universe and
most of the luminous matter we can observe is coupled to these fields.
The onset of star formation, the density and distribution of the
interstellar medium, the gaseous halos of radio galaxies and the
evolution of galaxies themselves are all controlled in large by the
action and presence of these magnetic fields. However, it is in
clusters of galaxies where a knowledge of the importance of evolution
of magnetic fields is crucial for our wider understanding of structure
formation and cosmology. The existence of cluster-wide magnetic fields
from diffuse synchrotron emission has been found in a number of
clusters (see \S \ref{halorelics}), with the radio emission spanning
several Mpc and in some cases following the filamentary networks of
galaxies. The magnetic field strengths derived from such emission
are intriguing as they suggest
field strengths which are dynamically significant, but not
dominant. As well as ordered fields in clusters, associated
measurements using the Faraday rotation of background or embedded
cluster radio sources have produced evidence for a tangled field,
although current constraints on the slope of the power spectrum are
poor. With improved observational constraints, clusters can provide an
excellent experimental environment in which to test theories of MHD
turbulence in large-scale structure and, with improved instrumental
characteristics and data processing, even weaker fields will become
observable, both within clusters of galaxies and in the wider web of
large-scale structure.

\begin{figure}[tbh!]
\footnotesize
 \begin{centering}
  \includegraphics[width=3.25in]{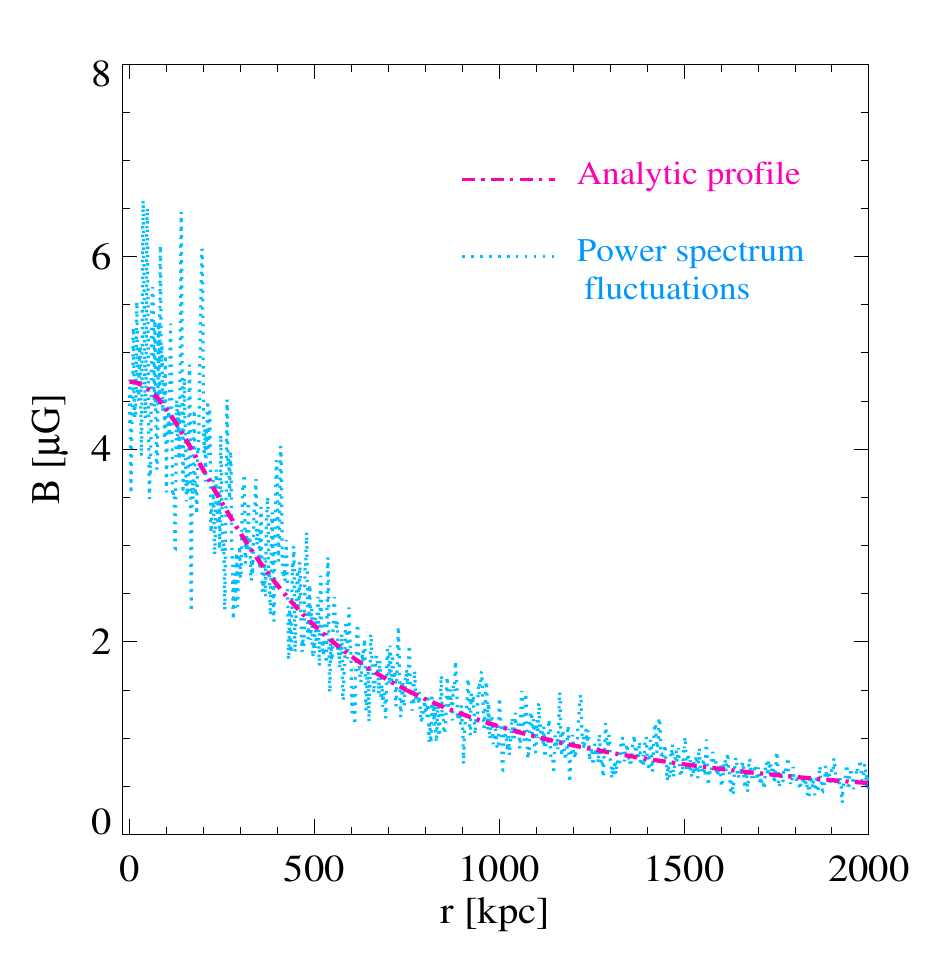}
   \caption{\small Radial profile of the magnetic field strength in the Coma cluster,
          inferred from VLA Faraday rotation measures.
	      Figure from \citet{Bonafede10}.
  }\label{fig:coma}
 \end{centering}
\end{figure}

Observations of such fields are important not only for cluster physics, but also 
for constraining the origin of cosmic magnetism more generally: early type seed 
fields may be truly primordial, with a seed field formed prior to recombination; 
alternatively fields can be produced by the Weibel instability - small scale plasma 
instabilities at structure formation shocks; late type fields can be injected into 
the WHIM via super-massive blackholes and other outflows. Within evolved clusters, 
highly efficient amplification is expected to occur; however, in spite of this, 
a clear distinction between the evolution of the magnetic field strength with redshift 
is expected in the case of early and late type fields.

In order to extract the maximal information contained in the polarization 
components of both background and embedded cluster radio sources it is necessary
to use the technique of RM Synthesis \citep{BB2005}, whereby the depolarization 
caused by Faraday rotation in wide bandwidth data when averaging for optimal 
sensitivity is avoided. This technique uses the pseudo-Fourier relation between
wavelength squared and Faraday depth to construct the Faraday depth spectrum along 
the line of sight for each direction, isolating features of particular Faraday 
depth and preserving the information content of the full bandwidth. The further 
analysis of such Faraday spectra is a field still in its infancy, although 
rapid progress is now being made, as these spectra provide a uniquely powerful tool 
for probing a wide range of different astrophysical environments.

Faraday rotation itself is caused when polarized emission passes
through a magnetized ionized medium, such as the ICM. The degree of
rotation which affects that emission is a linear function of both the
electron density and the parallel component of the magnetic field
strength along the line of sight. Unlike other methods commonly used
to probe the magnetic content of clusters it does not necessarily rely
on model dependent parameters, the presence of a non-thermal particle
population or assumptions of equipartition and provides independent
complementary information to field strength measurements derived from
such methods. The linear dependencies of rotation also make it ideal
for examining regions where both electron density and magnetic field
strength are expected to be low, such as the outskirts of clusters and
potentially the \emph{inter}-cluster medium and wider cosmic web of
large-scale structure.

Resolution in Faraday depth space is determined by $\Delta \lambda^2$, the width 
of the $\lambda^2$ coverage of the observation. A large $\Delta \lambda^2$ improves 
the resolution and also removes $n\pi$ ambiguities. 
RM synthesis with the VLA provides $\delta\phi_{\rm S} \approx 200$\,rad\,m$^{-2}$
(in S Band) or $\delta\phi_{\rm L} \approx 50$\,rad\,m$^{-2}$ (in L Band)
resolution in Faraday depth -- note this is the FWHM of the 
rotation measure transfer function (RMTF), not the \emph{accuracy} 
with which an RM can be recovered, which is a function of signal-to-noise. The maximum 
observable Faraday depth before bandwidth depolarization within a single channel 
becomes important is $>>5000$\,rad\,m$^{-2}$ at both L- and S-band, far higher
than is expected for cluster RMs. The maximum observable Faraday width of a 
feature which is extended in Faraday depth space (i.e. mixed rotation and emission) 
before strong depolarization occurs is $L_{\phi, max, L}\approx 35$\,rad\,m$^{-2}$;
$L_{\phi, max, S}\approx 140$\,rad\,m$^{-2}$.

Detailed studies of Faraday depth can give information on the magnetic
field distribution at different locations in galaxy clusters
\citep[e.g.][]{Bonafede10,Bonafede13}, in clusters in a different
physical state \citep[e.g.][]{Bonafede11} and in clusters at different
cosmological distance.

Recently \citet{Govoni13} demonstrated that simulated radio halos are
intrinsically polarized at full-resolution. The fractional
polarization at the cluster center is $\sim$ 15--35 \% with values varying
from cluster to cluster and increasing with the distance from the
cluster center. However, the polarized signal is undetectable if
observed with the comparatively shallow sensitivity and low resolution
of current radio interferometers.  However \citet{Govoni13} found
that surveys planned with the SKA precursors will be in principle be able
to detect the polarized emission in the most luminous halos known,
while the halos of intermediate and faint luminosity will still be
hardly detectable. In particular they showed that the VLA already has the
potential to detect polarized emission from strong radio
halos.


\subsection{Multi-wavelength Complementarity}\label{complements}

A number of ongoing and completed cluster surveys provide useful
catalogs for studying the non-thermal emission from the ICM, directly
and through stacked statistical results that will probe how the radio
properties scale with cluster mass and thermal properties.  
VLASS will be the top level radio
survey to study non-thermal properties in the Universe.

The primary energy band used in comparisons that constrain non-thermal
astrophysics of diffuse sources in galaxy clusters is the X-ray,
which probes the dynamical and thermal properties of the ICM.
Statistically, a large improvement in the number of known cluster diffuse radio
sources was obtained in a cross comparison by \citet{Giovannini99} 
of the NVSS radio survey with the the sample of X-ray-brightest Abell-type
clusters (XBACs; \cite{Ebeling96}). The XBAC sample comprises 283
clusters/subclusters from the catalogue of \citet{Abell89} (ACO)
detected in the ROSAT All Sky Survey (RASS).  
Indeed, most radio statistical analyses are derived from X-ray data
\citep{Schuecker01,Buote02,Cassano2010} and properties of radio
emission are characterized in general using X-ray temperature and luminosity
\citep[see e.g.][and references therein]{Giovannini02}. 
Predictions based on turbulent re-acceleration models agree well with
the radio observations of halos \citep{Cassano06}. In this sense,
strong spatial correlations between X-ray and radio brightness 
and between SZ-signal and radio brightness are also found in a number
of cases (e.g.\ \citep{Govoni01,Planck2013XXIX,Sommer2013}.

The upcoming eROSITA X-ray satellite\footnote{
\protect\url{http://www.mpe.mpg.de/eROSITA}} \citep{Predehl2010} will soon launch,
and will perform the first all-sky X-ray survey in over the two and half 
decades since the ROSAT All-Sky Survey (RASS).
Just as NVSS and RASS were complementary in many respects,
VLASS and eROSITA All-Sky Survey (EASS)  are well timed to complement each other. 
EASS is expected to find $\sim9\times10^4$ clusters 
above a mass of $0.7\times10^{14} M_{\odot}$ \citep{Pillepich2012}.  
Using the radio halo power -- X-ray luminosity
scaling relations of \cite{Cassano2013}, all clusters with an 
$L_{\rm{X}} \geq 10^{45}$~erg/s at $z<0.6$ will be detectable at $>$5-$\sigma$
for an L or S Band survey reaching 100~$\mu$Jy.
For a standard $\Lambda$CDM cosmology and a typical ``on'' fraction
of 30\%, this is roughly 80 clusters in a wide-area (30,000 deg$^2$) 
survey with VLASS, using the predicted cluster counts in \citep{Merloni2012}.
While uch lower radio powers will be accessible at lower redshifts ($z\lesssim0.2$, 
see Figure \ref{fig:scalings}), this argues that a deep survey or targeted 
follow-up of EASS-discovered clusters will be necessary to probe to lower 
radio powers (and $L_{\rm{X}}$) at high-$z$.

Optical information, such at that provided by the 
Sloan Digital Sky Survey (SDSS)\footnote{\protect\url{http://www.sdss.org/}},
Dark Energy Survey (DES)\footnote{\protect\url{http://www.darkenergysurvey.org/}}, 
or the Large Synoptic Survey Telescope (LSST)\footnote{\protect\url{http://www.lsst.org/}},
is another important way to investigate the dynamics of
cluster mergers \citep{Girardi02}. The spatial distribution and
kinematics of galaxy members allow us to detect substructures and to
analyse possible pre- and post-merging groups, and to distinguish
between evolving mergers and remnants. Moreover, optical data are
complementary to X-ray information because the ICM and galaxies react
on different time-scales during a collision \citet{Roettiger97}. 
The importance of combining X-ray and optical data to study merger scenarios 
has been clearly shown by, for example, the simulations of the Multi-wavelength 
Sample of Interacting Clusters (MUSIC) project \citep{Maurogordato11}.

Increasingly, microwave observations of the galaxy clusters are 
being used to probe galaxy cluster astrophysics as well.
The {\it Planck} satellite recently completed the first all-sky survey
to exploit the inverse Compton scattering of photons from the cosmic
microwave background (CMB) -- known as the
Sunyaev-Zel'dovich (SZ) effect -- to locate previously unknown galaxy
clusters.  Analysis of the first 15.5 months of Planck data has
produced a catalog of 1,227 clusters
\citep{Planck2013XXIX}. 
A large fraction of these clusters are massive, disturbed systems which 
have extended diffuse radio emission in the form of halos and/or relics.
ACTPol \citep{Niemack2010}, the polarization and detector upgrade to the Atacama Cosmology
Telescope, is now performing a deep, arcminute-resolution SZ survey of
roughly 4,000 square degrees of the southern sky at 
$\delta \geq -40^\circ$, accessible in a wide VLA survey of the sky.  
Projections for ACTPol indicate it will locate $>$1,000 clusters
with a higher median redshift than those in the {\it Planck} catalog.


\begin{figure*}[tbh!]
\footnotesize
 \begin{centering}
  \includegraphics[width=6in]{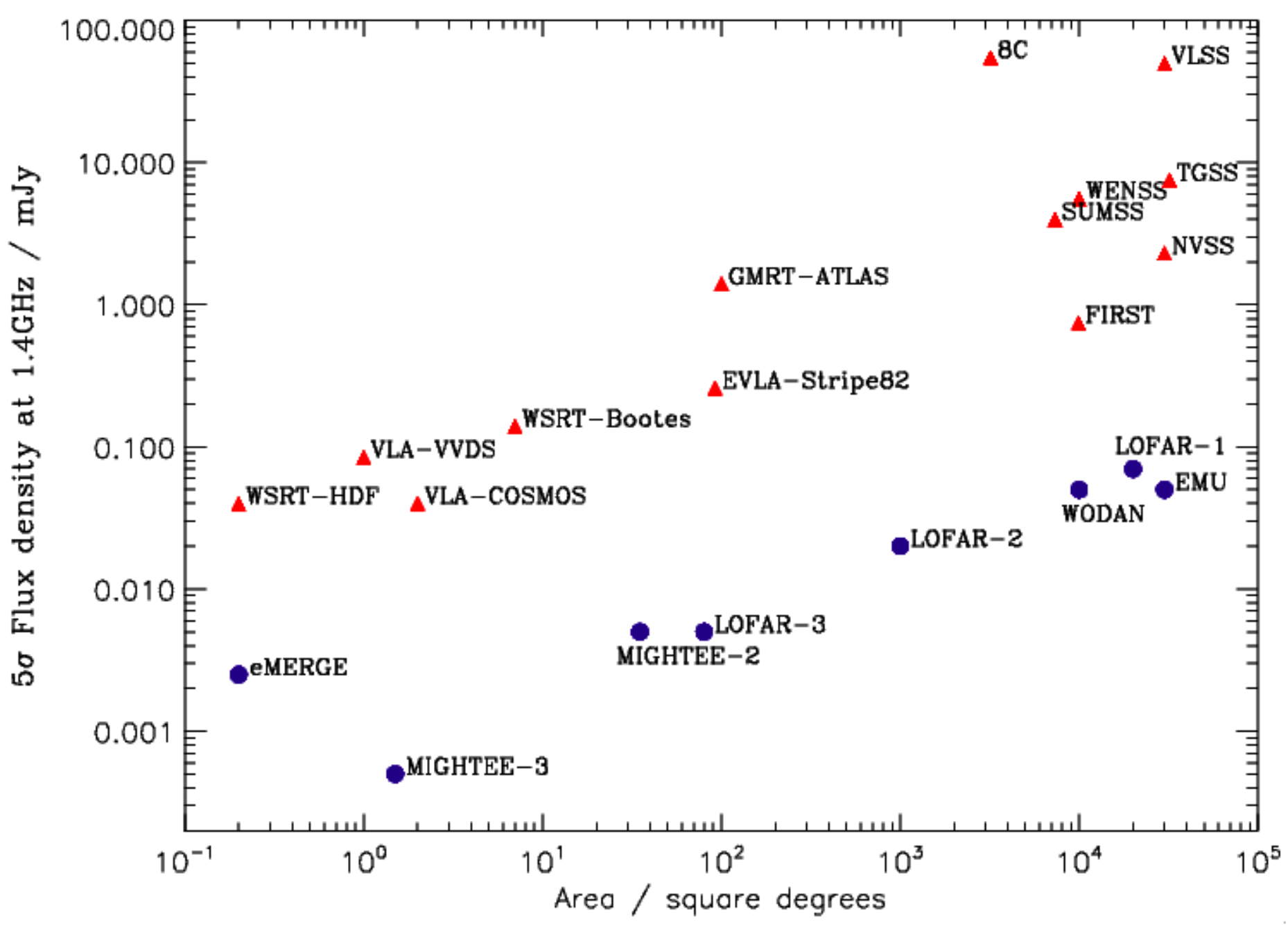}
  \caption{\small Summary of point source sensitivities and areas covered by radio 
	  surveys that are ongoing, completed, or planned.  
	  Figure is from Jarvis et al.\ 2014 \cite{Jarvis2014}.
  }\label{fig:rad_surveys}
 \end{centering}
\end{figure*}

\subsection{Concurrent and Upcoming Radio Surveys}\label{surveys}

A number of completed, ongoing, and planned radio surveys -- such as 
NVSS, FIRST, WENSS, WISH, VLSSr, POSSUM, WODAN, SUMSS, TGSS, EMU, MIGHTEE, 
and LOFAR-MSSS to name a few -- also will complement VLASS.  These
surveys will be summarized in a concurrent white paper by the 
VLASS Science Organizing Committee (Myers et al. 2014, in prep.).
We include here a figure from Jarvis et al.\ 2014 \cite{Jarvis2014} 
summarizing these surveys' sensitivities and areal coverage 
(see Figure \ref{fig:rad_surveys}), noting
the sensitivities listed are largely valid for unresolved, compact 
(point) sources, and do not address the sensitivities required for diffuse
extended emission.

After the discovery of the Coma radio halo at 408 MHz with the 250-ft
radio telescope at Jodrell Bank \citep{Large59}, confirmed from
interferometric Cambridge One-Mile telescope radio data
\citep{Willson70}, observations carried out mostly with single dish
radio telescopes, and the WSRT found about 10 other clusters with a
diffuse halo-type radio emission \citep[see][]{Hanisch82}.

A significant breakthrough in the study of radio halos and other
diffuse cluster radio sources was obtained thanks to the NRAO VLA Sky
Survey (NVSS).  \citet{Giovannini99}, combining radio data from the
NVSS and X-ray catalogues, detected 18 new halo and relic candidates,
in addition to the 11 already known, owing to the good surface
brightness sensitivity of the VLA D configuration used for NVSS. 
All new candidates were confirmed by more sensitive targeted follow-up 
observations, mostly performed with the VLA. \citet{Kempner01} presented seven new
candidates from a search in the Westerbork Northern Sky Survey
\citep[WENSS][]{Rengelink97} at 327 MHz.  Recently, other extensive
radio observations have been published, such as the recent Giant Metrewave
Radio Telescope (GMRT) survey of massive galaxy clusters at $z=0.2$--0.4
\citep{Venturi07,Venturi08}.

The NVSS is even now still a great resource for finding diffuse
cluster radio sources. Most of the halo and relic sources studied in
detail with deep multi-frequency observations started with a preliminary
identification within the NVSS. In the latest cluster diffuse emission
review \citep{Feretti12} identified the current sample of 42 radio
halos, 39 relics (including double relics), and 11 minihalos. This
number is increasing however as new more sensitive observations are
made.

In spite of the many encouraging results obtained, we do not know yet
a few important points for a deeper knowledge of the non-thermal
emission in clusters of galaxies.
Key is the occurrence and the luminosity function of diffuse
radio sources.  The correspondence between cluster mergers and the
presence of a diffuse radio emission is well established, and this
dichotomy is supported by many observational and theoretical
papers. Luminous relaxed X-ray clusters are not expected to show a
radio halo or relic.  What is uncertain still is if all bright X-ray
clusters in a strong merger phase have diffuse non-thermal
emission. The crucial question remains, namely how the collision of
massive clusters gives rise to the wide range of observed non-thermal
properties which run the gamut from radio quiet to the presence of a
radio halo and/or one or more radio relics. Since mergers with similar
global properties (e.g., mass, X-ray luminosity, radio-galaxy
luminosity function) often exhibit very different non-thermal
properties, the key to understanding the origin of the diffuse radio
emission is likely to lie in the details of the complex interactions
of the cluster constituents (dark matter, intra-cluster gas, galaxies)
during a merger event, and how they relate to and affect the
non-thermal components, i.e., relativistic particles and magnetic
fields embedded in the ICM.  In order to shed light on this point a
survey with a better sensitivity with respect to the NVSS will be
necessary.

Another crucial point is the redshift distribution of radio halos and
relics. This is an important component missing in studies of the
evolution of magnetic fields properties in clusters.  Most of the
clusters studied up to now are at $z < 0.3$. Only 9 radio
halos, 5 relics and 1 minihalo are known at $z > 0.3$. However the
recent detection of the radio relics and halo in the El Gordo massive
cluster at $z = 0.87$ \citep{Lindner13}, suggests that a possible
population of diffuse sources in high redshift clusters could be
present, but we need a deeper and higher angular resolution survey
with respect to NVSS.

A third point is the study of structure beyond galaxy clusters.
Cosmological theories and simulations predict that galaxy clusters
are connected by intergalactic filaments along which they accrete mass.
Shocks from infall into and along the filaments are expected
to accelerate particles. These accelerated particles can emit
synchrotron radiation if cosmic magnetic fields are present.
Attempts to detect diffuse radio emission beyond
clusters, i.e.\ in very rarefied regions of the intergalactic space,
have shown recent promise in imaging diffuse synchrotron radiation of
very low level.

Some examples of this emission are the filament surrounding the
cluster ZwCl2341.1+0000 \citep{Bagchi02,Giovannini10}, the bridge
connecting A399 and A401 \citep{Planck2013XXIX}, the X-ray filament
between A222 and A223 \citep{Werner08}, the structure in A3444
\citep{Giovannini2009}, and the diffuse radio bridges connecting the
halo to the relic observed in Coma
\cite{Kim1989,Giovannini90,BrownRudnick2011} and A2255
\cite{Govoni05,Pizzo08}.

All these data support the existence of an intergalactic
magnetic field more widespread and somewhat lower than that in the
intra-cluster medium within clusters.  This field may represent the
seed field for galaxies and clusters, and may play an important role
in the formation of large-scale structure.


\section{Observational requirements to address forefront galaxy cluster 
science}\label{requests}

In this section we consider the fact that the VLASS will be designed
to accommodate a wide range of scientific goals and thus currently the
frequency (or frequencies) as well as configuration(s) of the survey
are uncertain. We therefore consider combinations of frequencies and
VLA configurations and highlight the implications for extended cluster
emission studies.

\begin{table}[tbh]
\footnotesize
\begin{center}
\caption[]{\small Characteristic physical and angular scales of 
extended radio emission features in clusters at $z=0.1$--0.5}\label{table:features}
\begin{tabular}{lccccc}\\
\hline \noalign {\smallskip}
Feature & Scale & $\theta(z=0.1)$ & $\theta(z=0.2)$ & $\theta(z=0.3)$ & $\theta(z=0.5)$\\
	& (kpc)     & ('') & ('') & ('') & ('')\\
\hline \noalign {\smallskip}
AGN Jets   & 10--100    & 5.4--54  & 3.0--30  & 2.2--22  & 1.6--16  \\
WATs/NATs  & 10--100    & 5.4--54  & 3.0--30  & 2.2--22  & 1.6--16  \\
Relics     & 10--1000   & 5.4--540 & 3.0--300 & 2.2--220 & 1.6--160 \\
Minihalos  & 50--300    & 27--162  & 15--90   & 11--66   & 8.2--49  \\
Halos      & 1000-1500  & 540--810 & 300--450 & 220--340 & 160--250 \\
Bridges/Filament& $>$1500 & $>$810 & $>$450   & $>$340   & $>$250   \\
\hline \noalign {\smallskip}
\end{tabular}
\end{center}
\end{table}

In Table \ref{table:features} we summarize common cluster radio features and
their physical and angular scales for a few redshifts of interest.  
Note that because of effects of the volume of the observable Universe, merger 
rate, the
higher efficiency of cooling by inverse Compton scattering of CMB photons at
high redshifts, the cluster halo numbers are expected to peak in the redshift 
range $z=0.2$--0.5 \citep{Cassano06}. 
However the recent detection of 2 radio relics and a halo 
in the high-$z$ ($z = 0.87$) SZ-selected massive cluster ``El Gordo'' 
suggests that an exciting population of diffuse sources 
in high redshift clusters could exist \citep{Lindner13}.

\subsection{Angular and linear scales}\label{scales}

From the science drivers described above for this white paper, it is clear 
that the VLASS must be able to probe scales from tens of kpc
to $\gtrsim$1~Mpc over a wide range in redshifts. In particular:

\begin{itemize} 

\item Mpc scale halos and (potentially) large scale filamentary structure 
are the principle drivers for the largest angular scale (LAS) requirements.
For instance, a radio halo of 1 Mpc (typical value for this class of sources), 
has an angular extent of $\sim 9^{\prime}$ at $z=0.1$, $\sim 3^{\prime}$ at
$z=0.4$, and $\sim 2^{\prime}$ at $z=0.8$.

\item Minihalos are generally $\sim100-300$~kpc in scale, so are a less 
severe of a constraint except for the most nearby systems. For instance,
a 300 kpc minihalo would have a LAS$\sim6.5^{\prime}$ at $z=0.04$.

\item Radio relics can be more than one Mpc long in one direction but are 
compact and filamentary in the direction perpendicular to the shock that 
generates them. In a sky survey this is not an issue in terms of LAS since 
the areal coverage includes the entire relic and the narrow axis should be 
much less stringent of a constraint than halos or minihalos. 

\item The radio emission from individual galaxies does not pose severe
constraints on the largest angular scales, exception made for possible
Mpc--size nearby extended radio galaxies. On the other hand, angular
resolutions of the order of few arcseconds, i.e. in the range 
$3^{\prime\prime}-20^{\prime\prime}$ would be necessary to cover both the 
nuclear activity and to study the emission along the jets and lobes over
a wide range of redshifts.

\item In terms of angular resolution, a range between 
$10^{\prime\prime}-20^{\prime\prime}$ would be wide enough to accommodate 
observations of minihalos, giant radio halos, and radio relics over a wide 
range of redshifts and intrinsic linear scales.

\item For the detection of diffuse radio sources, the sensitivity to low 
surface brightness emission is the critical parameter. 
The surface brightness of cluster diffuse sources can be as low as
1 -- 0.2~$\mu$Jy arcsec$^{-2}$.   
Assuming a survey depth of 100~$\mu$Jy RMS (see Sect. 4.4), we will need
a relatively large beam to image these low brightness sources (S-Band, 
D array will produce images with a brightness sensitivity of 0.2 $\mu$Jy 
arcsec$^{-2}$ or even better; C array images in the S Band will have only 
a sensitivity of 2 $\mu$Jy arcsec$^{-2}$).
\end{itemize}

\subsection{Observing frequencies and polarization studies}\label{obsfreq}

Two important considerations regard the spectral and polarization studies 
of the emission. Both are critical to understanding the underlying 
astrophysics of these sources. 

Mpc--scale cluster radio sources, i.e. both classical GHz radio halos and 
relics, have a spectral index of roughly $\alpha \sim -1.3$. 
While relics can be studied at up to 5 GHz with 
sufficient sensitivity to smooth large scale emission 
(e.g. \cite{Giacintucci2008}), only the Bullet Cluster (\cite{Liang2000})
and the Coma cluster have been 
imaged above 1.4 GHz.
Radio minihalos have similar spectral indices and current studies of 
minihalo
systems show that they can be studied at up to, and possibly above, 5 GHz
(\cite{Giacintucci2014}).

For a few cases, the spectra of radio halos and relics are not well described 
by powerlaws, and a steepening at higher frequencies is observed. 
A clear example is the prototype radio halos in the Coma cluster 
\cite{Thierbach2003,Brunetti2013}. 
The location of the break is closely related to the (re)acceleration 
processes at play, and is thus important to constrain.

The 2--4 GHz frequency range (S Band) is best suited to detect the 
spectral break in nearby relics and minihalos; S Band observations
in D array be a critical configuration in which to obtain data. 
On the other hand, the steeper spectrum halos driven by smaller-mass mergers 
and those relics due to adiabatic compression of fossil radio plasma will 
however be below detection limits at frequencies above 400~MHz.
For these source, observations with the VLA's new P Band will be
necessary.
See the VLASS white paper on VLITE by Clarke et al.\ as a possible small 
effort addition to obtain 10 antenna narrow-band P band data simultaneously with VLASS.

Polarization studies at low frequencies suffer significantly from 
depolarization as well as large Faraday rotations (see Mao et al. 2013 VLASS 
white paper) whereas the highest frequencies are only sensitive to very high 
RMs through the densest media. Studies in the 2--4 GHz regime are well-suited 
to RM synthesis studies of the ICM, where RMs can range from
$\sim$10--1000~rad/m$^2$ (see \S \ref{pol}). 
We recommend that any survey taken in the cm range be done with full polarization 
capabilities in mind.


\subsection{Considerations on the frequency selection}\label{frequencies}

Following on the legacy of NVSS and FIRST \citep{Becker1994}, it is tempting 
to consider L Band, especially considering that L Band is now 1~GHz wide, thus
covering a frequency range which is particularly interesting for galaxy
 cluster science.  
However, the fractional contamination of L Band is large, leaving typically 
60\% unflagged bandwidth, and a wide L Band VLASS may offer only an incremental 
improvement over FIRST and NVSS.  With the upcoming EMU\footnote{
\protect\url{http://www.atnf.csiro.au/people/rnorris/emu/}}, POSSUM\footnote{
\protect\url{http://www.physics.usyd.edu.au/sifa/possum/}}, and MEERKAT\footnote{
\protect\url{http://public.ska.ac.za/meerkat/meerkat-large-survey-projects}} 
surveys of the sky observable with the Australian and South African 
SKA Pathfinders, it becomes increasing difficult to justify L Band except 
to provide a Northern complement that extends their coverage to the entire sky.

\begin{table}[tbh]
\footnotesize
\begin{center}
\caption[]{\small Partial table of VLA bands, usable bandwidths, and the 
survey speeds to reach a continuum sensitivity of 100~$\mu$Jy RMS, reproduced
from {\it Capabilities of the Jansky VLA for Sky Surveys} 
(\protect\url{https://science.nrao.edu/science/surveys/vlass/capabilities}).
Here $\tau_{\rm int}$ is the integration time, 
$\theta_{\rm PB}$ is the angular extend of the primary beam (field
of view in a single pointing), and $\theta_{\rm res}$ is the resolution
provided in B array.
Bands $>$12~GHz were omitted as their survey speeds are diminishingly
small and they are unable to capture the large scale cluster emission,
which typically falls off at higher frequencies.
}\label{table:vlass_capabilities}
\begin{tabular}{lccccccc}\\
\hline \noalign {\smallskip}
Band & Freq & Bandwidth & $\tau_{\rm int}$ & $\theta_{\rm PB}$ 
  & $\theta_{\rm res}$ & Mapping Speed & Scan Rate \\
     & (GHz)& (GHz)     & (s)              & (')               
  & ('')               & (deg$^2$/hr)  & (deg/m) \\
\hline \noalign {\smallskip}
P & 0.23--0.47 & 0.20 & 8553 & 122 & 24.0 & ~0.98 & 0.01 \\
L &    1--2    & 0.60 &  ~37 & ~30 & ~5.6 & 13.90 & 0.65 \\
S &    2--4    & 1.50 &  7.7 & ~15 & ~2.7 & 16.53 & 1.56 \\
C &    4--8    & 3.03 &  4.4 & 7.5 & ~1.3 & ~7.21 & 1.36 \\
X &    8--12   & 3.50 &  3.9 & 4.5 & 0.78 & ~2.96 & 0.93 \\
\hline \noalign {\smallskip}
\end{tabular}
\end{center}
\end{table}

It is natural then to consider an adjacent frequency band to complement the 
above surveys.  
Using the survey speeds reported in 
{\it Capabilities of the Jansky VLA for Sky Surveys}\footnote{
\protect\url{https://science.nrao.edu/science/surveys/vlass/capabilities}},
which we reproduce in Table \ref{table:vlass_capabilities}, S Band 
stands as the fastest band for reaching a survey depth of 100~$\mu$Jy RMS.
Furthermore, due to its large bandwidth, spectral index and clean rotation 
measures of the ICM can be performed directly using S Band, while its data 
would leverage spectral indices jointly using L Band from complementary surveys.

The uniform target sensitivity of 100~$\mu$Jy RMS in Table \ref{table:vlass_capabilities}
could be misleading, considering that nearly all diffuse emission is brighter at lower 
frequencies with a spectral index steeper than $\alpha < -1$.
A survey utilizing P Band would provide valuable spectral leverage on known radio source
and would not have to reach the same sensitivity 
level as an S or L Band survey to detect the same diffuse features.   
A P Band flux limit of $\sim$ 850 (400)~$\mu$Jy RMS is conservatively 
equivalent to an S (L) Band limit of 100~$\mu$Jy RMS, (assuming $\alpha = -1$).
With this relaxed sensitivity requirement, a P Band survey can also be performed
more quickly.  It would take roughly 1225 hours to reach a level of 
500~$\mu$Jy RMS in a wide survey of 30,000 deg$^2$, compared to the 1815 hours 
for S Band to reach 100~$\mu$Jy RMS.
The situation for P Band only improves when considering most spectra are 
steeper than
$\alpha = -1$, and many more sources with lower mass and hence radio power 
would be detectable in P Band.
While excellent for steep spectrum sources, P Band and lower frequencies do 
have a drawback in that they suffer from rapid depolarization.

We note that the availability of a survey at the S and P Band will
largely increase scientific possibilities also in different research fields,
and the potential for new discovery is likely greater than it is for L Band.

Based on our considerations in the earlier sections, frequencies above
4~GHz (C Band and higher) are not suited for cluster diffuse emission
science.  While a good probe for polarization and total intensity
measurements of small scale features like AGN cores and jets in the
inner regions of clusters, its smaller field of view (and hence
limited LAS) and slower survey speed means many steep spectrum sources
would be missed by C Band in any configuration.

\begin{table}[tbh]
\begin{center}
\caption[]{\small Scales recovered in various survey configurations, using the 
information provided by 
\url{https://science.nrao.edu/facilities/vla/docs/manuals/oss2014a/performance/resolution.}
Our recommendations for probing the large scale diffuse cluster emission are highlighted in bold.
Configurations and bands that lead to insufficient resolution or severely limited
largest angular scales (LASs) recovered are denoted in {\color{red}red}.
The confusion limit is provided by the VLA Exposure Calculator, currently available at 
\url{https://obs.vla.nrao.edu/expCalc/14A/evlaExpoCalc.jnlp}.
We find that the confusion level that could be reached will be negligible for all 
but the deepest surveys (which are specifically designed to approach the confusion
limit).  We also note that complementary higher resolution information (e.g. S Band observations
in B Configuration) are necessary for ICM weather (Section \ref{weather}), AGN feedback 
(Section \ref{agn}), and for constraining the flux contributions from compact sources 
when observing the large scale diffuse emission. We indicate this higher resolution requirement in italics in the table.
}\label{table:configs}
\begin{tabular}{lcccc}
\\
\hline \noalign {\smallskip}
Band    & Config. & LAS & Resolution & Confusion \\
(Freq.) &  & '' & ''  & ($\mu$Jy)\\
\hline \noalign {\smallskip}
P (230--470~MHz) & A     & \color{red}{155} & 5.6 & -- \\
\bf P (230--470~MHz) & \bf B & \bf 515        & \bf 18.5 & \bf 39 \\
P (230--470~MHz)                     & C & 4150       & \color{red}{60}  & 390 \\
\hline \noalign {\smallskip}
L (1--2~GHz) & B & \color{red}{120} & 4.3 & -- \\
\bf L (1--2~GHz) & \bf C & \bf 970 & \bf 14  & \bf 10.72 \\
L (1--2~GHz) & D & 970              & {\color{red}46}  & 107.2 \\
\hline \noalign {\smallskip}
{\it S (2--4~GHz)} & {\it B} & \color{red}{{\it 58}}  & {\it 2.1} & -- \\
S (2--4~GHz) & C & 490              & 7.0 & 1.37 \\
\bf S (2--4~GHz) & \bf D & \bf 490 & \bf 23  & \bf 13.7\\
\hline \noalign {\smallskip}
C (4--8~GHz) & B & \color{red}{29}  & 1.0 & -- \\
C (4--8~GHz) & C & 240              & 3.5 & 0.21 \\
C (4--8~GHz) & D & 240              & 12  & 2.11 \\
\hline \noalign {\smallskip}
\end{tabular}
\end{center}
\end{table}

\subsection{Configurations}\label{configurations}

We briefly summarize the merits of different configurations below and
in Table \ref{table:configs}.

\noindent\uline{L Band, D Config:} res$=46''$, LAS $=970''$, 
confusion=110~$\mu$Jy/bm:
on the positive side it is sensitive to
the largest halos and good for steeper spectrum sources. 
However, the insufficient resolution to separate radio galaxies from halos 
and relics 
at higher redshift is a major issue for this choice of frequency and array.
Moreover, it is difficult to motivate from a uniqueness perspective, as
NVSS covered L Band in D configuration.\smallskip

\noindent\uline{S Band, D Config:} res$=23''$, LAS $=490''$, confusion=13.7~$\mu$Jy/bm:
confusion is not significant, this frequency will be
sensitive to Mpc halos above $z\gtrsim0.1$, 
the resolution is marginally low but close
to sufficient to separate radio galaxies from halos and relics. 
Sources with the steepest spectra, as well as lower mass and less powerful 
sources, will be missed by a survey at these frequencies.  
However, it will probe the critical break frequency region for many relics and
the Faraday studies are of interest for background AGN probes as well as 
relics.\smallskip

\noindent\uline{C Band, D Config:} res$=12''$, LAS $=240''$, confusion not an 
issue. 
On the positive side the resolution is sufficient for
separation of radio galaxies from halos and relics. A major negative is that 
the VLASS will lose sensitivity to Mpc scale halos at $z\lesssim0.3$. 
Also this high  frequency will miss steep spectrum sources, weaker relics 
that have a significant break around 2-4 GHz,
and even miss details of nearby large relics. \smallskip

\noindent\uline{P Band, D Config:} This should generally be avoided, since 
such combination of frequency and array is seriously contaminated 
by RFI and quickly confusion limited.\smallskip

\noindent\uline{L Band, C Config:}: res $=14''$, LAS $=970''$, 
confusion=11~$\mu$Jy/bm. 
Still, sensitivities as above to the largest relics and also the steeper 
sources 
but now there is sufficient resolution
as well to separate the radio galaxies from the halos and relics. This configuration
would probe a new regime between the
NVSS and FIRST for 21~cm studies, and is good for polarization and RM studies.
And since the vast majority of results for cluster sources have relied on NVSS 
and FIRST, a new survey here would allow for direct comparison with previous
results.\smallskip

\noindent\uline{S Band, C Config:}: res=7'', LAS=490'', confusion is not 
a problem. 
As above, very good for a large variety of reason for the science case. 
WAT/NAT studies would benefit greatly from this band, as it provides
both spectral leverage and better resolution for high-$z$ sources.
The main drawback is that, because of the high angular resolution,
the surface brightness sensitivity to diffuse cluster sources and extended radio
galaxies will be low.  It is necessary to properly map extended low brightness 
regions to study physical properties of clusters on large scales 
(e.g. magnetic fields).\smallskip

\noindent\uline{C Band, C Config:}: res=3.5'', LAS=240'', confusion not an 
issue.
However the resolution is too high and the 
available LAS unsuitable for the regions of interest. The band is
not appropriate for halos and relics. Only a small part of the galaxy cluster 
science could be addressed by this and higher frequencies.\smallskip

\noindent\uline{P Band, B Config:} 
The only major consideration for B configuration for this white paper would 
be the new
P band. This is well suited to having a large FoV, sensitivity to steep 
spectrum
emission, angular resolution of 18.5'' (well suited to separate radio 
galaxies from halos
and relics), LAS=515'' so Mpc scale features above redshift of $z>0.1$ are
detectable. 
The survey speed for steep spectrum ($\alpha=-1.3$) sources is much faster 
than any 
of the other frequencies and this frequency would provide spectral 
information in an interesting regime. 
Polarization studies are not ideal at this frequency, although it is possible 
that for the lower rotation measures of the WHIM it would be critical. 
The resolution matches that of TGSS, which provides 
interesting complementary information, despite the different sensitivities.\smallskip

\noindent\uline{{\bf Final comment}}:
The C and D configurations nominally have similar sensitivities to LAS for 
full synthesis (see Table \ref{table:configs}), and have roughly comparable 
survey speeds.  Ostensibly then C configuration for S or L band, and B 
configuration for P band, could probe sufficiently large scales for halos 
and LSS, while still resolving features that are 10's of kpc in scale.  
However, C configuration is limited in the number of baselines it provides 
that actually probe extended emission, and an S band survey in C config 
would suffer severe limitations in its ability to recover large scale structure 
(as discussed in \S \ref{scales}).
Observations in B configuration in S or L band would resolve smaller $\lesssim$ 
10~kpc scales, but would perform even worse for LSS, failing to 
distinguish radio halos from minihalos. 
The high-resolution A configuration runs the risk of not even being able to 
constrain features such as jets, NATs, WATs, and head-tail galaxies, which can 
span a few to a tens of arcseconds; in addition to missing important science on 
galaxies and feedback in cluster environments, these features must be removed 
in order to accurately recover the large scale, diffuse halo and minihalo emission.

However, as discussed before, the sensitivity to low surface brightness 
emission is the most critical parameter in the study of diffuse cluster sources. 
Observations in S Band will require the VLA D configuration to obtain images of 
sources with a surface brightness $\lesssim0.2 \mu$Jy arcsec$^{-2}$; 
C configuration S Band observations with an RMS of 100~$\mu$Jy will only a 
surface brightness level $\sim2 \mu$Jy arcsec$^{-2}$.

\subsection{Sky Coverage}\label{coverage}

\noindent\uline{Wide and Shallow:}
While the obvious choice to provide a legacy archive comparable to, but deeper
than, FIRST and NVSS (which are still being mined for new science), it is not
necessarily the best choice for detailed studies of cluster astrophysics.  
Despite this, the sensitivity limits considered for VLASS will have a significant
impact on the study of non-thermal emission from galaxy clusters.
If, however, VLASS covers two different bandwidths (specifically,
S and P Bands), the resulting archive will have a scientific value never
obtained in previous surveys.  Another strategy would be to complement
upcoming southern radio surveys in L band (see \S \ref{surveys}), 
using the VLA to cover the $\sim1/3$ of the sky inaccessible from South 
Africa and Australia.\smallskip
 
\noindent \uline{Narrow and Deep:}
Many of the arguments above are applicable here, except the deep drilling fields would 
be ideal places to use the power of frequency coverage provided by the VLA to
significantly enhance the science. For example, S band D + L band C + P band B would 
be a powerful combination to go deep on supercluster fields, where one could easily
probe much of the cluster extended emission over a very interesting environment.
The addition of higher frequencies to such deep fields would provide the resolution 
for deep studies of star formation, AGN activity, and lensed background sources behind 
the cluster potentials.\smallskip

\noindent \uline{Targeted Sample:}
Targeted follow-up of well-defined samples from e.g.\ eROSITA, {\it Planck}, and/or
ACTPol would also enable much of the science discussed above.  However, eROSITA's sample
will not be available by the start of VLASS, and a targeted survey in general would
not provide the kind of legacy archive comparable to FIRST and NVSS.\smallskip

%
%

\noindent \uline{Hybrid Survey: Wide and Shallow + Narrow and Deep:}
Another possibility to maximize the return of a new survey on galaxy 
cluster science would be the combination of a wide field survey, combined
with a deep survey over a smaller portion of the sky. This deep survey
could be performed at a different frequencies and in different configuration. 
One possible combination could be L band with C array for the wide and shallow survey, 
and S band and C array for the deep survey. As an example, the deep 
field could be the North Galactic Cap, where SDSS ancillary data are available
and eROSITA coverage will be deeper.\smallskip

\noindent \uline{Multi-frequency + Multi-Configuration Survey}
A final possibility would be a wide survey covering $\sim$10,000 deg$^2$
in L, S, and C bands in B, C, and D arrays respectively.
This would reduce the demand from VLASS for one particular configuration,
and thus impact concurrent science with the VLA less.
Nearly continuous spectral coverage from 1--8~GHz would powerfully leverage spectra 
-- mainly of compact structures such as AGN due to the limited LASs of L Band + 
B Array and C Band + D Array.  It would also leverage rotation measures of 
magnetic fields.


\section{Conclusions}\label{conc}

For cluster-scale diffuse emission, the choices of S Band + D Configuration,
L Band + C Configuration, and P Band + B Configuration offer sufficient 
resolutions for constraining galactic interactions and feedback in
cluster environments, while still probing large scale structure
and the bulk cluster environment itself.
A VLA survey using L Band + C Configuration would complement and build
upon the results of both NVSS and FIRST, while also probing larger scales 
than upcoming southern radio surveys.
However, from a uniqueness perspective, VLASS will likely have more impact 
on cluster astrophysics if it were to target P and/or S Band (in B and D Config, 
respectively), both of which provide larger fractional bandwidths
than L Band.  The advantage of P Band is that less sensitivity would be required 
to probe fainter cluster sources, while S Band is better suited for Faraday 
rotation measure studies and the detection of not very steep spectrum 
sources (e.g. $\alpha\gtrsim-0.7$ radiogalaxies and radio loud AGNs).
Finally, we note that complementary observations at higher resolutions 
(e.g. the $\sim$ few arcsecond resolution from S Band in B Configuration)
are indispensable for the study of AGN and ICM weather in cluster environments,
and aid in constraining the flux contributions from compact sources
that contaminate measurements of the diffuse emission.

While the push to probe higher redshifts and lower mass limits strongly 
favors a narrow and deep (or even targeted) survey strategy, we note that a wide survey
covering roughly 1/4--2/3 of the sky will have significant scientific 
return, discovery potential, and archival value.


\bibliography{vlass_refs}

\end{document}